# On-chip Time-bin to Path Qubit Encoding Converter via Thin Film Lithium Niobate Photonics Chip

Xiaosong Ren, Zhanping Jin, Xiaotong Zou, Xiaole Zhang, Hao Li, Lixing You, Xue Feng, Fang Liu, Kaiyu Cui, Yidong Huang, and Wei Zhang

## Affiliations

**Frontier Science Center for Quantum Information, State Key Laboratory of Low-Dimensional Quantum Physics, Beijing National Research Center for Information Science and Technology (BNRist), Electronic Engineering Department, Tsinghua University, Beijing, China**

Xiaosong Ren, Zhanping Jin, Xiaotong Zou, Xiaole Zhang, Xue Feng, Fang Liu, Kaiyu Cui, Yidong Huang, Wei Zhang

**National Key Laboratory of Materials for Integrated Circuits, Shanghai Institute of Microsystem and Information Technology, Chinese Academy of Sciences, Shanghai 200050, China.**

Hao Li, Lixing You

**Beijing Academy of Quantum Information Sciences, Beijing 100193, China.**

Yidong Huang, Wei Zhang

## Contributions

Wei Zhang and Xiaosong Ren proposed the quantum photonic chip scheme, were responsible for the chip fabrication, and designed the experimental systems. Xiaosong Ren, Zhanping Jin, Xiaotong Zou, and Xiaole Zhang carried out the experiment, collected and analyzed the experimental data. Hao Li and Lixing You developed and maintained the SNSPDs used in the experiment. Wei Zhang and Xiaosong Ren wrote the manuscript. Xue Feng, Fang Liu, Kaiyu Cui, and Yidong Huang contributed to the revision of the manuscript. Wei Zhang and Yidong Huang supervised the whole project. All authors have approved the version of the manuscript.

## Corresponding author

Corresponding author: Yidong Huang (yidonghuang@tsinghua.edu.cn), Wei Zhang (zwei@tsinghua.edu.cn)

## Abstract

The development of quantum internet demands on-chip quantum processor nodes and interconnection between the nodes. Path-encoded photonic qubits are suitable for on-chip quantum information processors, while time-bin encoded ones are good at long-distance communication. It is necessary to develop an on-chip converter between the two encodings to satisfy the needs of the quantum internet. In this work, a quantum photonic circuit is proposed to convert time-bin-encoded photonic qubits to path-encoded ones via a thin-film lithium niobate high-speed optical switch and low-loss matched optical delay lines. The performance of the encoding converter is demonstrated by the experiment of time-bin to path encoding conversion on the fabricated sample chip. The converted path qubits have an average fidelity higher than 97%. The potential of the encoding converter on applications in quantum networks is demonstrated by the experiments of entanglement distribution and quantum key distribution. The results show that the on-chip encoding converter can serve as a foundational component in the future quantum internet, bridging the gap between quantum information transmission and on-chip processing based on photons.

## Introduction

The rapid development of quantum information technology is triggering a potential information technology revolution[1]. From quantum computing[2] and quantum communication[3] to quantum sensing[4], technologies based on photonics demonstrate tremendous potential to break through the limitations of traditional information systems. With the development and increasing maturity of photonic quantum technologies, such as quantum key distribution (QKD)[5,6], quantum teleportation[7-9], and on-chip quantum computing and simulation[10,11], single-node quantum systems no longer meet the demands of advancement on large-scale and networked quantum information applications. Hence, the vision of the quantum internet also guides the development of photonic quantum information technology[12,13], which aims to connect distributed quantum processors for quantum information sharing and cooperative processing[14].

Integrated quantum photonic chips, recognized for their compact structure, high stability, and strong scalability, serve as a crucial platform for building photonic quantum information systems[15,16]. Encoding quantum information in the path of on-chip photons allows for convenient manipulation of quantum states using on-chip interferometers based on integrated components. Therefore, on-chip photonic qubits are commonly implemented using path encoding. Based on this encoding scheme, numerous quantum computing and simulation works have been demonstrated on photonic chips, including the Reck scheme[17], quantum logic gates[18,19], and quantum transport simulation[20]. As the scale and functionality of photonic quantum information systems developed, the interconnection between quantum photonic chips through optical fiber networks is required. In this network scenario, path-encoded qubits are unsuitable for long-distance transmission. It is necessary to convert them to qubits encoded on other freedoms suitable for optical fiber transmission[21]. Time-bin encoded qubits are commonly used for long-distance quantum information transmission in optical fibers as a representative of flying qubits. Utilizing time-bin encoded qubits, long-distance quantum applications such as QKD[22,23], quantum teleportation[7,24], and quantum secure direct communication (QSDC)[25] have been realized, with the transmission distances even over hundreds of kilometers[24,26,27]. Traditionally, the generation and measurement of time-bin encoded qubits often relies on unbalanced Mach-Zehnder interferometers (UMZIs)[28]. The probabilistic path selection in the passive UMZIs will lead to photon loss and prevent multi-stage cascading. This would hinder the implementation of complex qubit processing functions on chips. To satisfy the requirements of information processing within distributed quantum photonic chips and information transmission between them in an optical fiber network, it is necessary to develop on-chip converters between flying time-bin encoded qubits and on-chip path-encoded qubits.

High-speed optical switches provide temporal processing capabilities for the manipulation of time-bin qubits. It has been shown in the experiments that processing time-bin encoded quantum photonic states based on discrete high-speed optical switches with fiber pigtails, such as time-bin entanglement generation[29] and time-bin quantum logic gates[30]. Recently, the development of the thin-film lithium niobate (TFLN) platform has enabled the integration of such high-speed optical switches on chips[31]. Based on the electro-optic effect of TFLN waveguides, various high-performance integrated optical modulators have been developed, with modulation bandwidths even reaching hundreds of gigahertz[32-35]. Furthermore, TFLN waveguides exhibit low propagation loss[36-38], enabling the fabrication of low-loss matched optical delay lines (MODLs) for time alignment. Recently, a time-bin entanglement measurement on a TFLN chip without the need for the time post-selection of the single photon detectors has been demonstrated[39], showing the potential of the TFLN platform for processing time-bin encoded quantum photonic states.

In this work, we propose and demonstrate that the TFLN platform could offer a promising solution for realizing a time-bin to path encoding converter: using high-speed optical switches for directional routing of different time-bin photons and employing low-loss MODLs for their temporal alignment. The on-chip converters are designed and fabricated on TFLN chips with other photonic circuits for the demonstration of their performance and potential applications. First, on the chip, arbitrary time-bin-encoded photonic qubits are prepared and converted into path-encoded qubits. The single-qubit state tomography of the converted path-encoded qubit is performed with a high state fidelity of more than 97%, demonstrating the function of the converter. Then, the time-bin entangled state is prepared on the chip. The two photons in the state are sent to two converters, respectively, over optical fibers of 12.4 km. Then the state is converted to a path-entangled state, which is demonstrated by two-photon interference. An experiment of the entanglement-based QKD is performed as an example of the applications of the proposed on-chip converters, showing that they could serve as foundational components in the future quantum internet systems, bridging the gap between long-distance transmission and on-chip processing of quantum information in photonic qubits.

## Results

### Principle of the qubit encoding converter

The quantum photonic state encoded by time-bin is illustrated in Fig. 1a.i, where quantum information is encoded in the intensity and relative phase of photons in two time bins separated by a time interval $\Delta t$. The qubit can be expressed as:

$$|\varphi\rangle = \alpha|t_0\rangle + \beta|t_1\rangle \quad (1)$$

This encoding scheme is widely used in long-distance quantum communication applications, since the phase of the two adjacent time-bin photons is not sensitive to the slow disturbances in long-distance fiber transmission. While the state encoded in paths is shown in Fig. 1a.ii, the quantum information is encoded by the paths the photon passes through. The qubit can be represented as:

$$|\varphi\rangle = \alpha|0\rangle + \beta|1\rangle \quad (2)$$

On-chip quantum photonic state processing usually employs path-encoded states, since arbitrary manipulations can be easily achieved using on-chip interferometers.

The proposed on-chip converter is shown in Fig. 1b, realizing the conversion from time-bin qubits to path qubits. This scheme consists of a high-speed electro-optical switch (EOS) and MODLs with a delay difference of $\Delta t$ between the two paths. The EOS directs the photons from the $t_0$ bin to the long path and the photons from the $t_1$ bin to the short path. As a result, after the respective delays experienced by the $t_0$ and $t_1$ time-bin photons, the photons become temporally aligned at the output paths of the MODLs. The photons become indistinguishable in time and are distinguishable by their paths. The phase difference introduced by the different paths can be compensated using a thermal-optic phase shifter (TOPS), achieving an accurate conversion from time-bin encoding to path encoding.

Assuming the half-wave voltage of the EOS is $V_\pi$, changing the modulation voltage ($V_m$) by $V_\pi$ would enable the EOS to switch the status. As shown in Fig. 1c, a square wave modulation with a steep changing edge (less than $\Delta t$) transitioning from $V_0$ to $V_0 + V_\pi$ is used to switch the EOS from the through (all to long path) to the cross state (all to short path). In this scheme, the switching speed is critical. The electro-optic effect of TFLN photonic chips enables ultra-high-speed optical switching, providing the foundation for the implementation of this scheme.

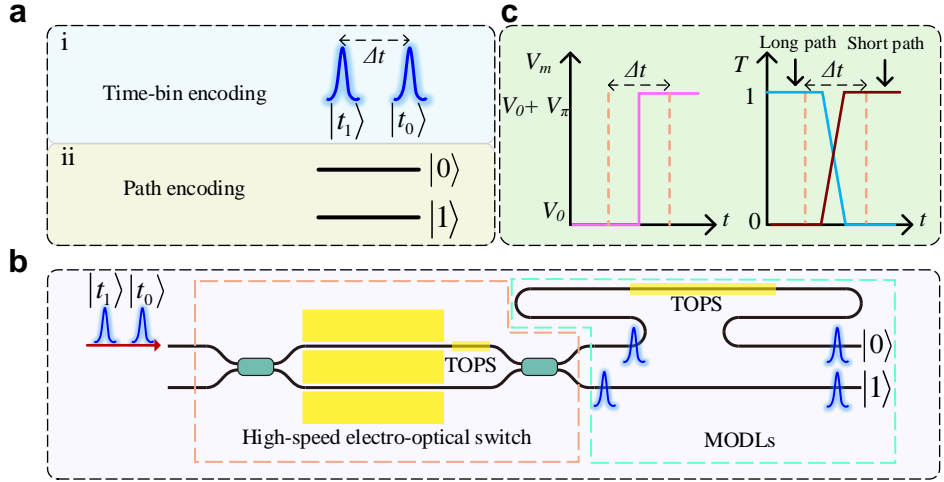

**Fig. 1**
**a**. The sketches of the time-bin and path encoded photonic qubits. **b.** The chip structure of the proposed qubit encoding converter, including a high-speed EOS and MODLs. **c.** The modulation signal ($V_m$) applied to the EOS, and the corresponding transmission ($T$) of the long or short path.

## Characterization of the qubit encoding converter

To characterize the qubit encoding converter, an integrated TFLN chip is designed and fabricated, as illustrated in Fig. 2. The chip consists of two sections. The first section, shown in Fig. 2a, is used to prepare arbitrary time-bin-encoded qubits. It includes a thermal-optic Mach–Zehnder interferometer (TO-MZI) to adjust the beam splitting ratio (BSR), which determines the relative intensity of the photons in the early and late time bins. Additionally, there is a MODL with a path length difference corresponding to a delay of 100 ps. This MODL is equipped with a TOPS to control the relative phase between the photons in the two time bins. These two components enable the realization of an arbitrary time-bin-encoded qubit, as described by Eq. (1), for any values of $\alpha$ and $\beta$. The second section, shown in Fig. 2b, comprises the time-bin to path encoding converter and a subsequent TO-MZI for characterizing the converted path-encoded qubit. The TO-MZI enables the complete tomography of the path-encoded state. A detailed description of the path-encoded qubit tomography procedure is provided in the **Methods** section. The fabricated TFLN photonic chip is shown in Fig. 2c. Its monolithic design and fabrication ensure excellent delay matching accuracy. After the electrical and optical packaging of the chip, the experiment is carried out.

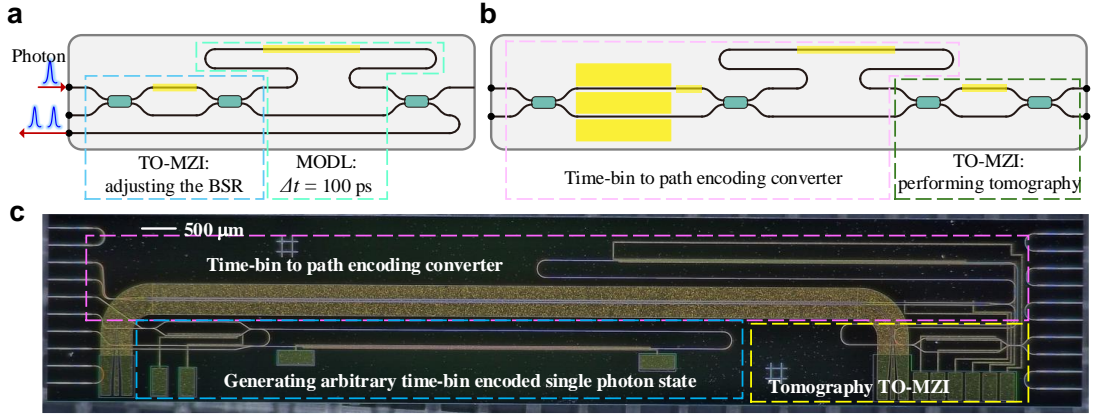

**Fig. 2**

The chip for characterizing the qubit encoding converter. **a.** The photonic circuit design for preparing an arbitrary time-bin encoded qubit. **b.** The photonic circuit design for the time-bin to path encoding converter followed by a TO-MZI for tomography. **c.** The optical microscope image of the fabricated TFLN chip.

The experimental setup for characterizing the encoding converter is illustrated in Fig. 3a. A pulsed laser emits optical pulses at a repetition rate of 100 MHz, and these pulses are used to pump a quantum light source based on spontaneous four-wave mixing (SFWM). As shown in Fig. 3b, in the quantum light source, the pulsed pump light is filtered by a dense wavelength division multiplexer (DWDM) with a 100 GHz bandwidth and a center wavelength of 1545.32 nm, corresponding to the International Telecommunication Union (ITU) channel of C40. The filtered pump light is then amplified by an erbium-doped fiber amplifier (EDFA). A long-pass filter is subsequently used to remove any leaked light outside the C-band. To stabilize the output power of the EDFA, a Proportional-Integral-Differential (PID) feedback system consisting of a 99:1 fiber coupler (FC), an optical power meter (OPM), and a variable optical attenuator (VOA) is used. Following the PID system, another DWDM filter system of C40 is used to remove the amplified spontaneous emission (ASE) introduced by the EDFA. A packaged silicon photonic chip containing silicon waveguides is pumped to generate photon pairs via SFWM. The photon pairs are then filtered using DWDMs with 100 GHz bandwidths and center wavelengths of 1537.40 nm (ITU channel C50) and 1553.33 nm (ITU channel C30) to separate the signal and idler photons, respectively. The signal photons are detected by a superconducting nanowire single-photon detector (SNSPD, PHOTEC Inc.) as heralding signals, while the idler photons serve as heralded single photons and are input into the time-bin qubit preparation section for qubit generation.

The time-bin-encoded qubit passes through an optical delay line (ODL) and a fiber polarization controller (FPC) before being coupled into the time-bin-to-path encoding converter and subsequent tomography measurement. Simultaneously, the pulsed laser provides a sync signal to trigger an arbitrary waveform generator (AWG), which outputs a square wave signal. The signal is amplified by a radio frequency (RF) driver and used to modulate the EOS on the TFLN chip. Synchronization between the modulation signal and the photon arrival time is achieved by adjusting the delay of the ODL. Additionally, all of the TOPSs are controlled via direct current (DC) ports connected to DC power supplies. After carefully calibrating each TOPS and precisely aligning the photon arrival time, the experiment of the conversion from time-bin-encoded qubits to path-encoded ones is carried out.

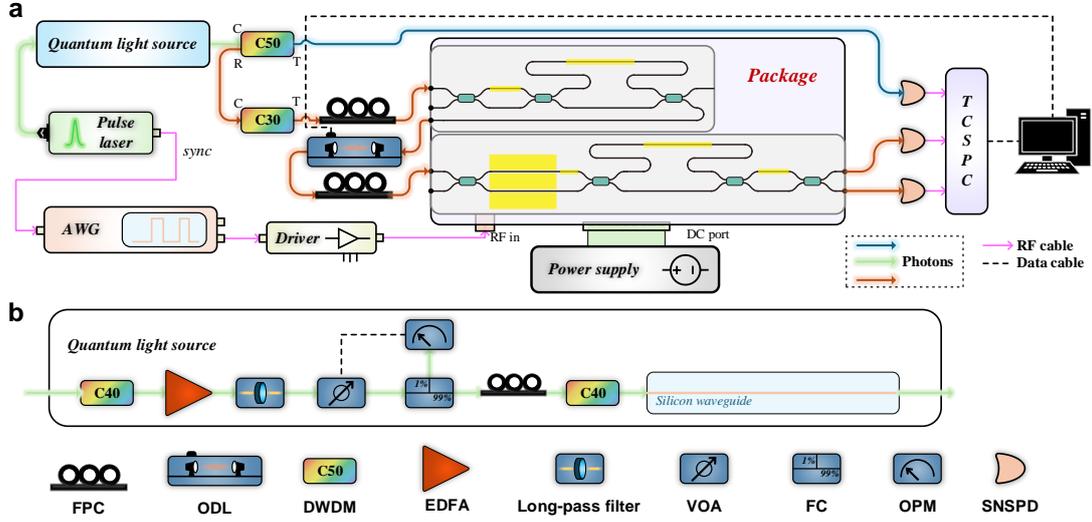

**Fig. 3**

**a.** The experimental setup for characterizing the qubit encoding converter. The quantum light source is pumped by a pulsed laser and generates the photon pairs, one of the photons in a pair is detected to provide the heralding signal, and the other serves as a heralded single photon. The heralded photon is encoded as a time-bin qubit. Then, the qubit encoding conversion and tomography are carried out. The pulsed laser and the AWG are synchronized, and the output of the AWG is amplified by an RF driver to modulate the EOS of the converter. **b.** The setup of the quantum light source for generating the photon pairs. SFWM spontaneous four-wave mixing, AWG arbitrary waveform generator, TSCPC time-correlated single photon counter, RF radio frequency, DC direct current, FPC fiber polarization controller, ODL optical delay line, DWDM dense wavelength division multiplexer, EDFA erbium-doped fiber amplifier, VOA variable optical attenuator, FC fiber coupler, OPM optical power meter, SNSPD superconducting nanowire single photon detector.

The experimental results are shown in Fig. 4. Fig. 4a shows the density matrices of two typical qubits. For six specific time-bin-encoded single-photon states, the fidelities of the reconstructed density matrices obtained from the converted path-state tomography all exceed 96%, with an average fidelity higher than 97%, illustrated in Fig. 4b. These results demonstrate that the proposed scheme can effectively convert time-bin-encoded qubits into path-encoded qubits. The complete tomography results for all six single-photon states are provided in the **Supplementary Information**. It is worth noting that the time jitter of the SNSPDs used is approximately 150 ps, while the time-bin separation $\Delta t$ is 100 ps. Hence, we could not measure the state tomography of the input time bin qubits. The converter effectively eliminates the dependency of the measurement of time-bin qubits on the time jitter of the SPDs, and the state tomography of the converted path qubits can be measured. Although these results include the imperfections in the preparation of time-bin qubits, the high fidelities of different converted states fully demonstrate the good performance of the converter.

It is worth noting that the time-bin to path converter can also convert path-encoded qubits to time-bin encoded ones by using the converter in reverse. We do not demonstrate the reverse conversion in this work since the time difference of time-bin qubits used in this work is 100 ps, which is comparable with the time jitters of the SNSPDs. Hence, it is difficult to measure the time-bin qubits directly in our experimental conditions. However, the performance of the reverse conversion could be estimated by the experiment results of the time bin to path conversion according to the principle of optical reversibility.

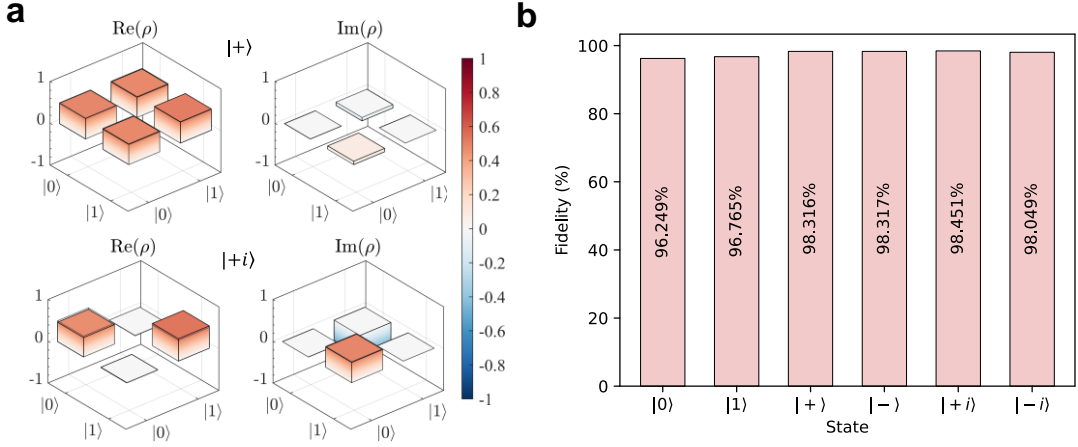

**Fig. 4**
**a.** Density matrices of the converted path encoded qubits of $|+\rangle$ and $|+i\rangle$. **b.** The fidelity of six certain converted qubits with an average fidelity higher than 97%.

## Application of the converters on entanglement distribution and QKD

To demonstrate the application of the proposed converter in quantum network, the experiments of entanglement distribution and QKD are carried out. The scheme of entanglement distribution is shown in Fig. 5a. The pulsed light passes through a UMZI, by which each optical pulse is converted to two pulses. Then, it is used to pump the quantum light source, which generates photon pairs in a time-bin entangled state. The signal and idler photons of the photon pairs are sent to users named Alice and Bob and measured, respectively. Alice and Bob can use the distributed entangled states to realize quantum information applications, such as QKD.

A TFLN chip is designed and fabricated for this experiment. On the chip, a photonic circuit identical to Fig. 2a is designed, where the BSR of the TO-MZI is set to 50:50 and the MODL has a delay difference of 100 ps. This part essentially functions as a UMZI with a 100-ps delay difference, converting an optical pulse of the pump light into two pulses, as illustrated in Fig. 5b. Figure 5c shows the photonic circuit of one user (Alice or Bob) for entanglement distribution and QKD. It includes a time-bin to path encoding converter. After the photon is converted into a path-encoded state, measurement basis selection and projective measurement in the Z- and X-bases are performed via three beam splitters (BS1, BS2, and BS3). Output 0 and output 3 are used to take the projective measurement of the Z-basis. Output 1 and output 2 are used to take the projective measurement of the X-basis, with the interference in BS3. The optical microscope image of the fabricated chip is shown in Fig. 5d. The chip integrates two photonic circuits for entanglement distribution measurement and QKD, corresponding to Alice and Bob, respectively.

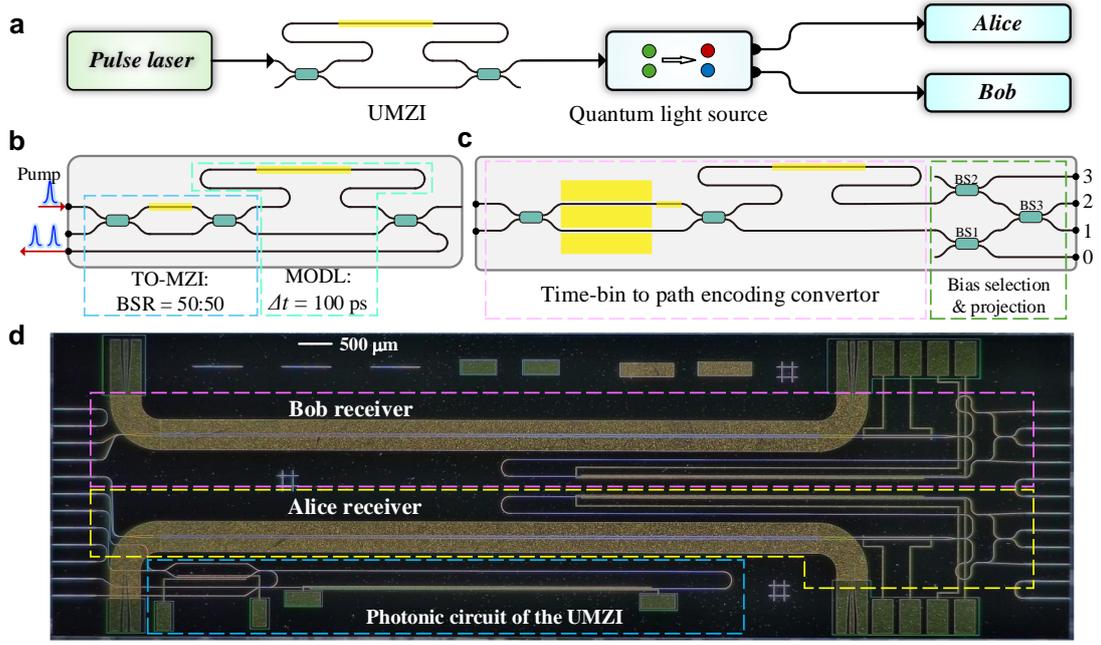

**Fig. 5**

**a.** The experiment design of time-bin entanglement distribution and QKD. **b.** The photonic circuit of the UMZI converting an optical pulse of the pump light into two pulses on the chip. **c.** The photonic circuit for one user of entanglement distribution and QKD, including a time-bin to path encoding converter and beam splitters for basis selection and projective measurement. Output 0 and output 3 are used to take the projective measurement of the Z-basis. Output 1 and output 2 are used to take the projective measurement of the X-basis. **d.** The optical microscope image of the fabricated TFLN chip.

The experimental setup for entanglement distribution and QKD is shown in Fig. 6. The pulsed light is filtered by a DWDM of channel C40, and its polarization is adjusted by an FPC before being coupled into the TFLN chip. The on-chip UMZI converts each pulse of the pump light into two pulses, which are used to pump the quantum light source based on SFWM, as shown in Fig. 3b. The quantum light source generates photon pairs in the entangled state:

$$|\Psi\rangle = \frac{1}{\sqrt{2}}\left(|t_{0,s}t_{0,i}\rangle + e^{i2\theta}|t_{1,s}t_{1,i}\rangle\right), \tag{3}$$

where the phase $\theta$ can be adjusted via the TOPS in the MODL. Subsequently, the signal and idler photons are separated by C50 and C30 DWDMs, and then propagate through dispersion-shifted fibers (DSFs) of 6.2 km in the laboratory, respectively. Finally, after passing through ODLs and FPCs, the photons are coupled into the photonic circuits of Alice and Bob, respectively. The photons output from the four output ports of Alice (Bob) are filtered by C30 (C50) DWDMs before they are detected by SNSPDs. Simultaneously, the pulsed laser and the AWG are synchronized via a sync signal. The square wave signals generated by the AWG are amplified by two RF drivers and applied to the RF ports of Alice and Bob, modulating their TFLN EOSs. Additionally, all TOPSs on the chip are controlled via DC ports connected to DC power supplies. After carefully calibrating each TOPS and precisely controlling the photon arrival times of Alice and Bob, the experiments of entanglement distribution and QKD are carried out.

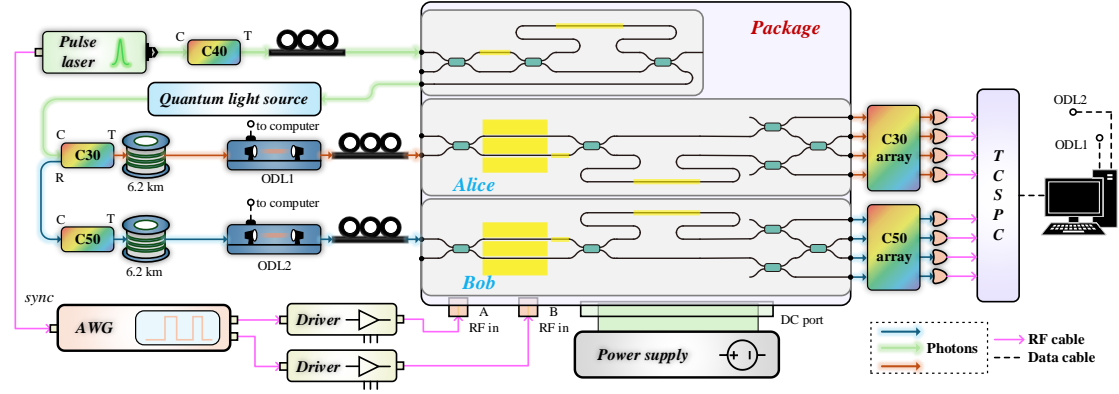

**Fig. 6**

Experimental setup for entanglement distribution and QKD. The pulsed pump light passes through the on-chip UMZI, then pumps the quantum light source to generate time-bin entangled photon pairs via SFWM. The signal and idler photons of the photon pairs are distributed to Alice and Bob through 6.2 km-long DSFs, respectively. Before the photons are injected into the photonic circuits, ODLs are used to align the photon arrival times with the square wave signals for the TFLN EOSs at both sides. The photons output from the four output ports of Alice (Bob) are filtered by C30 (C50) DWDMs before they are detected by SNSPDs. The pulsed pump light and the AWG are synchronized. The AWG outputs two square waves, and two RF drivers amplify the square waves to modulate the EOSs on the TFLN chip.

The experiment of entanglement distribution is made under both the back-to-back (without fiber transmission) condition and fiber transmission of 12.4 km. First, photon pairs of a time bin entangled state with a specific $\theta$ in Eq.(3) are prepared. The signal and idler photons are sent to the quantum photonic circuits at Alice and Bob, respectively. After they pass through the on-chip encoding converters, the time bin entangled state is converted to a path entangled state. To characterize the entanglement, two-photon quantum interferences of path entanglement are measured. The single photon detection events at the output ports of BS3 (Output 1 and 2) in both Alice and Bob are recorded under the scanning heating power of the TOPS in Alice's MODL when that of Bob's MODL is fixed at two values, which support interference phases of 0 and $\pi/2$ at Bob, respectively. The measured fringes of the two-photon interferences are shown in Fig. 7a and Fig. 7b. It can be seen that in the back-to-back case, the visibilities of the fringes are 91.34% ± 0.57% and 90.73% ± 0.61% under the two interference phases at Bob. After fiber transmission, the visibilities are 88.02% ± 1.16% and 90.79% ± 1.03%, respectively. All the values are higher than the criterion for violating the Bell inequality, showing that the encoding converters at the two sides can convert the freedom of the entanglement from time bin encoding to path encoding, maintaining good entanglement properties. To show that the time-bin to path encoding converters could handle different entanglement states, the phase $\theta$ of the source is scanned, and the signal and idler photons are recorded under a fixed measurement base of Alice and Bob. The results of the two-photon interference are shown in Fig. 7c and Fig. 7d. It can be seen that the visibilities of the fringes all exceed 90% under the back-to-back case and fiber transmission of 12.4 km. These results further confirm that the encoding converters are universal for time-bin encoding photons. It is worth noting that in these experiments, the fringe visibilities under fiber transmission decrease slightly than those under the back-to-back case. It is mainly due to the fluctuation of photons' arrival times at both sides when they propagate through the optical fiber.

By fixing the interference phases of Alice and Bob to support the X basis (Output 1 and 2) and using Output 0 and 3 as the Z basis, the entanglement-based QKD is implemented using the BBM92 protocol[40]. In the back-to-back case, the raw key rate (RKR) of the QKD system is approximately 331 bps with a quantum bit error rate (QBER) of about 4.36%. After the fiber transmission of 12.4 km, the RKR decreases to approximately 113 bps with a QBER of about 6.18%. The reduction in RKR is partly due to the transmission loss of optical fibers, and partly due to the polarization variation when photons propagate through optical fibers. The increase in QBER could be expected according to the decreases in fringe visibilities under fiber transmission, which is mainly due to the fluctuation of photons' arrival times.

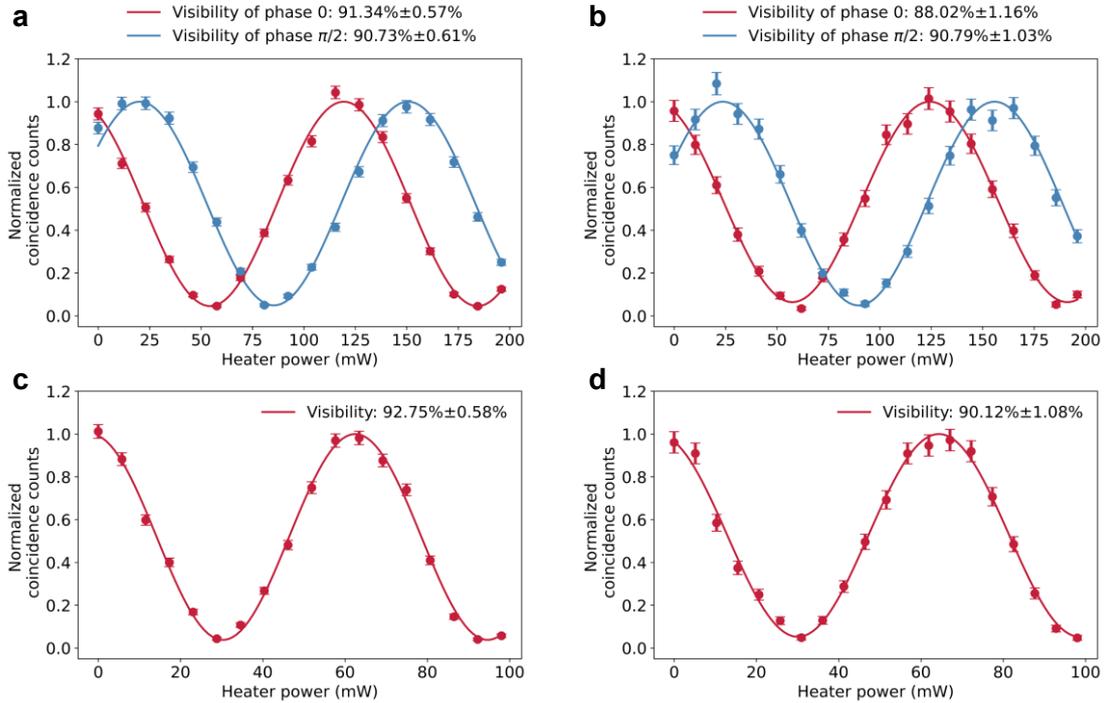

**Fig. 7**

Experiment results of entanglement distribution. The two-photon interference fringes of the signal and idler photons at the quantum photonic circuits of Alice and Bob with the path entangled state, measured under **a.** the back-to-back condition, and **b.** with fiber transmission of 12.4 km. The two-photon interference fringes when the phase $\theta$ of the source is scanned and the signal and idler photons are recorded under a fixed measurement base of Alice and Bob under **c.** the back-to-back case, and **d.** with fiber transmission of 12.4 km.

## Discussion

The fringe visibility of quantum interference in the experiment of entanglement distribution has room to improve. The impacts on the visibility include: (1) The limited extinction ratio (ER) of the EOSs. Driving by the square wave signals, the maximum ERs for the through and cross conditions of the EOSs are approximately 17 dB in the experiment. It leads to incomplete separation of photons from the two time bins into their paths, reducing the interference visibility. (2) According to the calibration process described in **Supplementary Information**, when the EOS is set to the state as a 50:50 beam splitter, the UMZI structure at Alice's side, composed of the EOS, MODL, and BS3, has an interference ER of ~12 dB, while that at Bob's side has an interference ER of more than 20 dB. It shows that the performance of the quantum photonic circuit at Alice's side is lower than that at Bob's side. These impacts are mainly due to the deviation in the fabrication process of the quantum photonic circuits, which could be addressed by improving the fabrication accuracy and stability.

The raw key rate of the QKD system could be further improved. After packaging, the insertion loss of Alice's photonic circuit is approximately 10 dB, while Bob's has a higher insertion loss of 15.5 dB due to packaging issues. It can be expected that the raw key rate could be improved by better packaging processing. Furthermore, the quantum light source used in the experiment is based on SFWM in a silicon waveguide. To achieve a high coincidence-to-accidental ratio (CAR), the intensity of the pump light is set at a relatively low level, leading to a low entangled photon pair generation rate of approximately 60 kHz. If a quantum light source with much higher brightness is used, a significantly higher raw key rate could be expected. Especially, periodically poled TFLN waveguides could be used to generate entangled photon pairs via spontaneous parametric down-conversion (SPDC) based on the second-order nonlinear effect[41]. It has the potential to be fabricated at the same TFLN chip together with the qubit encoding converters and provide bright photon pairs with high CAR.

The fiber transmission distance in the experiment of entanglement distribution is 12.4 km. It is worth noting that a longer entanglement distribution distance could be expected if the photon arrival times and the square wave signals for the EOSs could be synchronized at both sides. Since photon arrival times would fluctuate when they propagate along actual long-distance fibers, the square wave signals need to be adjusted accordingly to track these fluctuations. The synchronization can be achieved by two methods: (1) Active feedback stabilization systems should be employed to stabilize photon arrival times at both sides[7]. With this method, only an initial calibration between the modulation signals and photon arrival times is required. (2) Optical pulses for the synchronization should co-propagate with the entangled photon pairs through the same fiber[42]. In this method, the synchronous pulses experience the same time fluctuations as the photons. Using these synchronous pulses to trigger the square wave signals at the two sides, the synchronization could be achieved.

In this work, we have demonstrated the application of the time-bin to path encoding converter specifically in quantum communication applications of entanglement distribution and QKD. It is worth noting that the architecture is a universal structure for time-bin-to-path encoding conversion, applicable to any scenario requiring the transduction of time-bin encoded states into path-encoded states. For instance, in future photon-based quantum internet or distributed quantum computing systems, time-bin encoded photons can serve as flying qubits to transmit quantum information, for example, in quantum teleportation. When these flying qubits arrive at certain nodes, the proposed converter can transduce them into path-encoded on-chip qubits. This enables subsequent on-chip quantum computing or quantum information processing tasks.

# Methods

## Fabrication and packaging of the TFLN chip

The TFLN photonic chip is fabricated on an x-cut lithium-niobate-on-insulator (LNOI) wafer with a top lithium niobate (LN) layer thickness of 600 nm. First, a 200 nm chromium (Cr) layer is deposited via electron beam evaporation (EBE) onto the top LN layer. Electron-beam lithography (EBL) and inductively coupled plasma (ICP) etching are used to pattern the Cr layer, defining it as a hard mask for the subsequent LN etching. Then, LN with a thickness of 300 nm is etched via ICP, and the Cr mask is removed, thereby defining optical waveguides, MZIs, ODLs, BSs, and so on. Next, an 800 nm layer of amorphous silicon (a-Si) is sputtered. EBL and ICP etching are employed to pattern the a-Si, which serves as a second hard mask for etching the LN to form edge couplers. Another 300 nm of LN is then etched via ICP, followed by removal of the a-Si mask, resulting in a two-layer LN edge coupler structure. A 1 μm thick silicon oxide ($SiO_2$) buffer layer is deposited via plasma-enhanced chemical vapor deposition (PECVD). Ultraviolet (UV) lithography and reactive ion etching (RIE) are used to open windows in the $SiO_2$ buffer layer, preparing for the deposition of traveling-wave electrodes on LN. Subsequently, UV lithography and a lift-off process are utilized to pattern a 200 nm thick nickel-chromium (NiCr) alloy, forming the heating resistors for TOPSs. Gold electrodes with 1000 nm thickness are then fabricated using UV lithography and lift-off. These form the traveling-wave electrodes for LN modulation and provide electrical connections to the heating resistors. Finally, a 2.8 μm thick $SiO_2$ cladding layer is deposited via PECVD. UV lithography and RIE are used to pattern this cladding, creating the necessary $SiO_2$ cover for the edge couplers. Windows are also opened in the cladding layer at the electrode pad regions. The chip facets are ultimately diced and polished to achieve optimal edge-coupling efficiency.

The fabricated chip is then packaged within a box that provides optical and electrical interfaces. The chip is mounted onto a custom-designed heat sink. A miniature thermistor is attached to the heat sink to monitor the chip temperature, and a thermoelectric cooler (TEC) is placed beneath the heat sink for active temperature control. The electrical connections are made by wire-bonding the pads of the on-chip traveling-wave electrodes and TOPSs electrodes to a printed circuit board (PCB) inside the package. These are routed to external RF and DC ports. The RF ports consist of 65 GHz-bandwidth GPPO connectors, while the DC port is a multi-channel pin connector. Fiber arrays are aligned and attached to the chip's edge couplers using UV-curable adhesive, enabling optical input and output.

## Single Qubit Tomography

To perform single-qubit tomography, it is necessary to experimentally obtain projection measurement results under three different measurement bases. For a single-qubit state $|\varphi\rangle$, its density matrix $\rho$ can be reconstructed through quantum state tomography. Generally, the density matrix can be expressed as[43]:

$$\rho = \frac{1}{2}\sum_{i=0}^{3} S_i \sigma_i,$$
$$S_i = \langle \varphi | \sigma_i | \varphi \rangle,$$
$$\sigma_0 = \begin{pmatrix} 1 & 0 \\ 0 & 1 \end{pmatrix}, \sigma_1 = \begin{pmatrix} 0 & 1 \\ 1 & 0 \end{pmatrix}, \sigma_2 = \begin{pmatrix} 0 & -i \\ i & 0 \end{pmatrix}, \sigma_3 = \begin{pmatrix} 1 & 0 \\ 0 & -1 \end{pmatrix}$$

(4)

where $S_i$ is the Stokes parameters, which can be calculated by the projection measurement results. To ensure the physical validity of the density matrix—namely, its non-negative definiteness and Hermitian properties—the maximum likelihood method has been proposed for reconstructing the

density matrix of a quantum state[44]:

$$\rho = \frac{T^\dagger T}{\text{Tr}(T^\dagger T)}, T = \begin{pmatrix} t_0 & 0 \\ t_2 + it_3 & t_1 \end{pmatrix},$$

$$L(\rho) = \sum_{i=1}^{6} \frac{[N\langle \varphi_i | \rho | \varphi_i \rangle - n_i]^2}{2N\langle \varphi_i | \rho | \varphi_i \rangle}, \quad (5)$$

$$\rho_{rec} = \arg\min(L(\rho))$$

where $N$ is the total photon number at one measurement, $n_i$ is a certain experimental projective measurement result, and while $\langle \varphi_i | \rho | \varphi_i \rangle$ is the ideal projective measurement. The reconstructed density matrix $\rho_{rec}$ is obtained by solving the maximum likelihood optimization problem.

In our experiment, the projective measurements are performed by configuring the tomography TO-MZI to either the through state or the state of the BSR of 50:50, and setting the phase of the TOPS in MODL to 0 or π/2, respectively, as detailed in Table 1.

**Table 1.** The configuration of the TO-MZI and the TOPS under different projective measurements

| Projective basis | Z | X | Y |
|---|---|---|---|
| BSR of the Tomography TO-MZI | through | 50:50 | 50:50 |
| Phase of the TOPS in MODL | 0 | 0 | π/2 |

The fidelity of the reconstructed density matrix can be calculated using:

$$F = \left[ \text{Tr}\left( \sqrt{\sqrt{\rho_{aim}} \rho \sqrt{\rho_{aim}}} \right) \right]^2 \quad (6)$$

where $\rho_{aim}$ is the ideal density matrix for a certain path-encoded qubit.

# Data Availability

All data needed to evaluate the conclusions in the paper are present in the paper and/or the Supplementary Information. Additional data related to this paper may be requested from the authors.

## Acknowledgements

This work was supported by the Quantum Science and Technology-National Science and Technology Major Project (No. 2024ZD0302502, WZ), National Natural Science Foundation of China (92365210, WZ), Tsinghua Initiative Scientific Research Program (WZ), and the project of Tsinghua University-Zhuhai Huafa Industrial Share Company Joint Institute for Architecture Optoelectronic Technologies (JIAOT, YH).

## Supplementary Information